\documentclass[12pt,preprint]{aastex}

\newcommand{\be}{\begin{equation}}
\newcommand{\ee}{\end{equation}}
\newcommand{\bea}{\begin{eqnarray}}
\newcommand{\eea}{\end{eqnarray}}

\begin{document}


\title{Hard radio spectra from reconnection regions in galactic nuclei}


\author{G.T. Birk\altaffilmark{1}, A. R. Crusius--W\"atzel\altaffilmark{1} and H. Lesch\altaffilmark{1}}
\affil{Institut f{\"u}r Astronomie und Astrophysik der Universit{\"at}
M{\"u}nchen, D-81679 M{\"u}nchen, Germany}

\altaffiltext{1}{also: Centre for Interdisciplinary Plasma Science (CIPS),
Garching, Germany}


\begin{abstract}
Extremely flat and inverted radio spectra as observed in galactic nuclei
and BL Lac sources are still a challenge for fast particle acceleration models.
Continuous acceleration by electric fields in reconnection regions can result
in almost constant particle distributions and thereby in inverted
synchrotron spectra independent from the details of the injected spectrum. 
These spectra are calculated from a spatially averaged
diffusion transport equation in momentum space that includes  
systematic momentum gains and losses as well as finite particle escape 
lifetimes.
\end{abstract}


\keywords{Galaxies: nuclei -- BL Lacaertae objects: general --
Radio continuum: galaxies -- Acceleration of particles}


\section{Introduction}
Flat and even inverted radio spectra, 
$I_\nu\sim \nu^{-\alpha}$ with $\alpha <0$, are
a common feature of the non-thermal emission of galactic nuclei.
For example Sgr A$^{*}$, the very center of the Milky Way,  as well as the
galactic center arc region \citep{lesch92, dusch94}
exhibit an inverted spectrum with $\alpha \ge -1/3$.
The same holds for the nucleus of M81 \citep{reu96}.
\citet{slee94} found that compact radio continuum
cores in about  70 \% of radio emitting elliptical and S0 galaxies
have a flat or inverted spectrum with a median spectral index of $\alpha=-0.3$.
Observations of the Seyfert galaxy NGC 1068 indicate an inverted spectrum
with index $\alpha=-0.3$, as well \citep{mux96}. 
For active galactic nuclei (AGN) radio surveys by \citet{kue81} and  
\citet{ray00} have accumulated a huge wealth of data about flat radio spectra.
Of special interests are the flat radio spectra of the BL Lac sources 
Mrk 421 \citep{tos98}  and Mrk 501 \citep{edw00},
both also strong TeV-emitters. However,
AGNs and BL Lacs exhibit relatively steep
power-law spectra at frequencies higher than the radio regime. 
The complete 
electromagnetic spectrum of these objects is a result of different
energy loss and gain mechanisms. Consequently, the total spectra show 
flat and steep power-law spectra up to the radiation cut-off
\citep{mil96, ter99}.
Flat spectrum radio sources are known to be variable over the entire 
accessible part of the electromagnetic spectrum down to the shortest 
timescales probed so far \citep{wag98}.

Flat spectrum radio spectra have been attributed
to self absorption. For the self-absorption peak frequency to extend  over some
decades specific geometrical properties in the source are required - 
this is well known as "cosmic conspiracy" \citep{marsch77, cot80}. 
This model depends on the source size as $I_\nu \propto\nu r$, where 
$r$ is the radius of a conical geometry.
It implies that the linear size of the source is inversely
proportional to the emission frequency, 
which is not supported by interferometric data \citep{cot80}.
Of special interest is the galactic center region, in which it was clearly
proven that its flat radio spectrum is not due to self absorption 
\citep{lesch88, yus89}.

The evidence for flat, optically thin radio spectra in several active 
galactic nuclei has 
been presented by \citet{hugh89, mel96, wang97}. All these authors consider
different Fermi-like acceleration schemes (either multiple shocks or 
second order Fermi mechanism) to be responsible for the hardness
of the electron energy spectra.

In the present contribution it is suggested that optically thin 
synchrotron emission due to 
hard electron spectra produced in magnetic reconnection regions
can explain the nature of the flat/inverted spectrum radio sources. 


\section{Particle acceleration in reconnection regions}

Any acceleration model of charged particles must ultimately
be based on the energy gain of the particles in electric fields. However,
the dynamics of the respective plasma processes involved
can be quite different. The principal acceleration scenarios
(reviews are given, e.g., by \citet{sch86, kir94, kui96})
may be divided in
the Fermi I and Fermi II mechanisms, shock-drift acceleration
(which in fact can be considered as an example of Fermi I acceleration),
diffusive shock acceleration, plasma wave acceleration,
electrostatic double layers and the energization of particles in
reconnection events. In contrast to the other 'classic' mechanisms
it was only more recently that reconnection was discussed as a process
that does not only convert magnetic field energy
to plasma bulk motion and heating but also plays an important
role in the context of fast particle acceleration \citep{sch91,
vek95, lesch98, sch98, lit97, lit99}.

In the context of shock acceleration much elaborated
work has been done concerning the shape and evolution
of the energy distribution of the accelerated particles (e.g. \citet{kui96, 
ach00} and ref. therein).
One may note that in the diffusive shock scenarios the
injection problem appears,
i.e. one has to assume a high-energy electrons to start with.
Within many applications only the modification of power law spectra
by shocks which are
assumed for the injected particle population in the first place are
investigated.
What is more, very flat spectra are very hard to explain in the context of 
simple shock acceleration \citep{mel96}.

Nothing comparable to the detailed studies on the energy spectra
of shock accelerated particles has been done for
the reconnection scenario. In this contribution we analyze the energy spectra
that can be expected to be caused by high-energy particles accelerated
in dc electric fields in
reconnection regions starting from the space-averaged transport equation for
the one-particle distribution function that represents the high-energy
particle component.

In three-dimensional configurations with $B\ne 0$ everywhere
a finite electric field component parallel to the magnetic field $E_s$
is necessary for the onset of magnetic reconnection \citep{hes88, sch91}.
Ideal Ohm's law is to be violated
\be
{\bf E} + \frac{1}{c}{\bf v}\times{\bf B} = {\bf R}\label{1} \ne 0
\ee
where $\bf R$ is some yet unspecified non-idealness.
In the associated generalized electric potential
\citep{sch88, sch91}
\be
U=-\int E_sds=-\int R_sds\label{2},
\ee
where the integral is evaluated along the magnetic field lines that
penetrate the reconnection region, charged particles can be efficiently
accelerated. It should be mentioned that in general $U$ is not an
electrostatic potential.
Note that particles are not necessarily accelerated up to
energies associated with $U$ but expression (\ref{2}) gives an upper limit
of the energy level of the accelerated particle population.
The electric potentials  are often associated with non-ideal flows like
steady rotational motion or (non-)stationary shear flows.
The nature of the non-idealness $\bf R$ and the
reconnection process determines the strength, structure and time
dependence of the electric potential and thereby the particle energization.
One can think of different kinds of non-idealness in Ohm's law that 
allow for the reconnection process to start.
Since astrophysical plasmas are on large spatial scales highly collisionless
the Coulomb resistivity is negligible. Nevertheless, Ohm's law can
locally be violated by $R_s=\eta j_s$, where
$j_s$ is the magnetic field aligned component of the current density,
provided that some anomalous resistivity
$\eta$ is caused by microturbulent stochastic electromagnetic fields
(e.g. \citet{hub85} and ref. therein). The microturbulence can be caused by
currents. When the current density exceeds a critical value, unstable
kinetic waves can be excited that in their non-linear evolution result
in stochastic microturbulent electromagnetic fields. 
The same concept $R_s \sim j_s$ for
$j>j_{crit}$ holds for stochastic ion acoustic double layers as a source
for anomalous dissipation \citep{hae93}.
In relativistic plasmas, in particular,
finite particle inertia can be responsible for
the non-idealness $R_s\sim \partial j_s/\partial t + \nabla \cdot(\bf v \bf j
+ \bf j \bf v)$ as was recently discussed in the context of extragalactic
jets \citep{lesch98}.

In the present study we are rather interested in the resulting
power-law indices of the energy spectra
of the high-energy particles than in their maximum energy.
The spectra result from the combined effects of direct acceleration of
charged particles in the dc electric fields in reconnection regions and
radiative loss processes. For the acceleration the yet unspecified quantity
$R_s$ plays the crucial role.
In the simplest standard scenario of two-dimensional stationary reconnection
a constant electric field perpendicular to the plane where a
projection of one magnetic field component changes sign occurs, e.g.,
\citep{par63, bis93}. This electric field is a result of the non-idealness
$R_s$ that is assumed to be constant in the reconnection region.
Let us now consider a configuration with
a dominant constant magnetic field component parallel to the
direction of the associated electric current.
The electric field parallel to this magnetic field component
is approximately $E_s$ and therefore the fast particle acceleration is 
constant in time. A finite pitch angle of the accelerated particles allows 
for the radiative losses. 
To be more specific let us consider a generic shear
configuration. If a plasma flow, say ${\bf v}_s=v_s(x,z){\bf e}_y$ 
is applied to an homogeneous magnetic field ${\bf B}=B_z{\bf e}_z$ a
shear component forms $B_y = \int B_z\partial v_s/\partial zdt$ that results, 
in particular, in an electric current flowing parallel to $B_z$.
Given some localized dissipation 
(caused, e.g., by a supercritical current density) magnetic reconnection can
take place. In a stationary state a constant \citep{bis93} 
inflow ${\bf v}_i=v_i{\bf e}_x$
towards the reconnection region occurs and the magnetic shear
is converted to heat and kinetic energy as fast as it is built
up by the external shear flow $v_s$. 
The convergent velocity $v_i$ 
can be expressed as a function of the local Alfv\'en
velocity and the (constant) reconnection rate \citep{bis93}.
The quasi-stationary magnitude of the electric field component that 
causes the acceleration of fast particles
along $B_z$ can then be estimated without specifying for the 
non-idealness $R_s$ as it is possible in a more elaborated
general three-dimensional reconnection model \citep{sch88, sch91}. 
In our example, we find
\be
E_z\approx\frac{1}{\Delta_{{\rm rec}}c}\int_{\Delta_{{\rm rec}}}dz\int_Tdt
v_i\frac{\partial B_y}{\partial t}\label{0}
\ee
where $\Delta_{{\rm rec}}$ and $T$ denote the extent of the reconnection 
region along $B_z$ and the typical time scale of shear formation.

In what follows, we describe the particle acceleration by a 
an average constant electric field. 
In particular, the acceleration force does not depend on the 
momentum of the charged particles.

\section{The momentum balance equation for the accelerated particle 
population}

The population of high-energy particles that do not interact 
with each other can be described by
the one-body distribution function $f({\bf r}, {\bf p}, t)$
that obeys the Liouville equation
\be
\frac{\partial f}{\partial t} + \frac{d {\bf r}}{d t} \nabla_{\bf r} f
+ \frac{d {\bf p}}{d t} \nabla_{\bf p} f =  0
\label{3}.
\ee
Though this equation contains the full dynamical information it is of 
limited use since this non-linear differential equation in seven variables
is usually unsolvable for physical situations of interest.  
Fortunately, we need not to know the full information of the particle dynamics
in a rigorous manner. First, will assume
pitch angle isotropy. This idealization is motivated by the fact that in 
reconnection regions  Alfv\'enic and magnetosonic wave fields with relatively
high-energy densities are to be expected which cause an efficient particle 
scattering in momentum space \citep{sch86}. Additionally, self-excited 
Alfv\'en turbulence maintain quasi-isotropy, if the systematic velocity
of the high-energy particles exceeds the Alfv\'en speed of the ambient plasma
\citep{kui96}.
Second, when asking about the energy spectrum of the accelerated 
particles in a specified astrophysical object the detailed spatial 
dependence of the distribution function is often of no interest. 
Rather it is usually sufficient to concentrate on a 
distribution function space-averaged  over the volume $V$. Thus, we deal with
the isotropic distribution
$f^\prime(p,t)=\int d {\bf r} f({\bf r}, p, t)/V $. The 
total particle number is calculated as $N(t)=4\pi\int f^\prime p^2dp$.
The loss in information is not too high a price to pay in order to obtain a 
transport equation in momentum space one can really deal with.
The continuous momentum gains and losses $\Delta p$ the particles experience
can be described by a momentum operator
\be
{\cal L}_p = \frac{1}{p^2}\frac{\partial}{\partial p}[(\dot p_{gain}+
\dot p_{loss})p^2]\label{5},
\ee
if one concentrates, as in the present study, on systematic temporal
momentum gains $\dot p_{gain}$ and losses $\dot p_{loss}$.
The finite characteristic time $\tau$ a particle stays in the acceleration 
region gives rise to a further sink in the balance equation for
$f^\prime(p,t)$. In the reconnection scenario particles can escape 
$E_s\ne0$-regions by following reconnected field-lines outward the non-ideal
regions. Some explicit source term $S(p)$, on the other hand, 
represents the injected particle population.
Taking these effects into account the temporal evolution of the 
particle distribution is governed by the equation
\citep{jon70, sch84, sch86}
\be
\frac{\partial f^\prime (p, t)}{\partial t} + {\cal L}_p  f^\prime (p, t) + 
\frac{f^\prime(p, t)}{\tau} = S(p)\label{4}
\ee
which holds, if the distribution function $f({\bf r}, p, t)$ is separable in 
the space and the momentum variables, and if the momentum operator
${\cal L}_p$ is space independent.

It should be noted that equations(\ref{5}) and (\ref{4}) hold only for 
momentum changes $\Delta p \ll p$ during a time interval 
$\Delta t \ll \tau$. If one wants to include the effects of stochastic 
diffusion in momentum space, which may become important, e.g.,
for a microturbulent resistivity, a Fokker-Planck  diffusion term could be 
incorporated  \citep{sch84, sch86} in equation(\ref{5}).

The momentum change terms in equation(\ref{5}) in our case
can be specified as follows. The dominant loss processes
that are mainly responsible for the non-thermal radiative phenomena 
are synchrotron and inverse Compton radiation.
We will not consider Bremsstrahlung ($\dot p_{loss} \sim p$) here,
since this effect usually is dominated by the other ones mentioned.
Both processes scale as  \citep{lon81}
$\dot p_{loss}=-kp^2$ with $k=\frac{4}{3}\sigma_{\rm T}c^2U$ where
$\sigma_{\rm T}$ is the Thomson cross-section and $U$
is the magnetic field or photon energy density, respectively.
The momentum gain of a single charged particle in the acceleration region is 
given by
\be
\dot p_{gain}=\frac{d}{dt}(\gamma m v) = q E_s = q R_s\label{6} 
\ee
where $\gamma$ is the Lorentz factor and $q$
is the electrical charge.
In this contribution we exclusively consider 
the case $q R_s= \zeta=$const. (see discussion in section~2)
and thus, solve the transport equation
\be
\frac{\partial f}{\partial t}+ 
 \left(\frac{2\zeta}{ p}-4kp \right)f + 
\left(\zeta-kp^2\right)\frac{\partial f}{\partial p}+
\frac{f}{\tau} = S(p)\label{7}
\ee
where the primes are now omitted.
Equations of this kind can, in general, be solved by means of the inverse
Laplace transformation \citep{mel80a}.

\section{Power law particle distributions}

First, we solve the steady-state version of 
equation(\ref{7}) for different source populations.
The resulting distribution function reads
\be
f(p)=f_0p^{-2}\vert\zeta -kp^2\vert^{-1}
\left\vert\frac{1+p\sqrt{k/\zeta}}{1-p\sqrt{k/\zeta}}
\right\vert^{-\frac{1}{2\tau\sqrt{k\zeta}}}
\label{8}\ee
where $f_0$ depends on the injected source population by the integral
\be
f_0=\int^p{p^{\prime 2}
\left\vert\frac{1+p^\prime\sqrt{k/\zeta}}{1-p^\prime\sqrt{k/\zeta}}
\right\vert^{\frac{1}{2\tau\sqrt{k\zeta}}}S\left(p^\prime\right)
dp^\prime}
\label{8a}.
\ee
If the acceleration is negligible as compared 
to radiative losses, the stationary energy spectrum 
equation(\ref{8}) shows the pure aging behavior (see \cite{lon81}) caused
by synchrotron or inverse Compton losses 
$f(p)\sim p^{-4}$ where $f(p)dp$ is the particle number in the 
momentum interval $dp$. If the losses are negligible, the spectrum is 
given by $f(p)\sim p^{-2}e^{-p/\zeta\tau}$.

We note that the distribution function given by equation(\ref{8})
may diverge at the momentum $p=p_c=\sqrt{\zeta/k}$ where gains and
losses balance. When it is rewritten as
\be
f(p)=f_0k\left(\frac{p}{p_c}\right)^{-2}
\left(1+\frac{p}{p_c}\right)^{-\frac{1}{2\tau\sqrt{k\zeta}}-1}
\left\vert 1-\frac{p}{p_c}\right\vert^{\frac{1}{2\tau\sqrt{k\zeta}}-1}
\label{9}\ee
it becomes clear that the function is well behaved for
$2\tau\sqrt{k\zeta}< 1$, with $f(p_c)=0$.
A pile-up of particles occurs if
$2\tau\sqrt{k\zeta}> 1$. This is equivalent to
$\tau>\sqrt{t_{acc}t_{loss}}/2$ if one defines the acceleration and
loss timescale as $\dot p_{gain}=p/t_{acc}=\zeta$ and
$\dot p_{loss}=-p/t_{loss}=-kp^2$,
respectively. This means that the particles are collected near $p=p_c$
when the escape time (which depends on the geometry of the acceleration region)
is longer than half the geometric mean of the timescales for acceleration and
radiative losses. Otherwise the particles escape before they can
accumulate.

It needs mentioning that
the pile-up at the critical momentum $p_c=\sqrt{\zeta/k}$ would be
smoothed out, if a Fokker-Planck diffusion term was incorporated in 
equation(\ref{7}). This was done by \citet{sch84} for the case of
shock acceleration.

We now turn to determine $f_0$ in equation({\ref8})
for different injection spectra.
For a monoenergetic source population 
$S(p)=s_0\delta(p-p_0)$ we find
\be
f_0=s_0p_0^2\Theta(p-p_0)
\left(\frac{1+p_0\sqrt{k/\zeta}}{1-p_0\sqrt{k/\zeta}}
\right)^{\frac{1}{2\tau\sqrt{k\zeta}}}
\label{10}\ee
where $\Theta$ denotes the Heaviside function. In the next section we will
calculate spectra for this source population.
For a relativistic Maxwellian source population
$S(p)=s_0 {\rm exp}(-ap)$, where $a$ is a constant, one obtains 
\be
f_0=s_0\int^p{p^{\prime 2}
e^{-ap^\prime}\left\vert\frac{1+p^\prime\sqrt{k/\zeta}}{1-p^\prime
\sqrt{k/\zeta}}
\right\vert^{\frac{1}{2\tau\sqrt{k\zeta}}}
dp^\prime}
\label{11}\ee
where $c$ is some integration constant.
In the limit of negligible particle escape, i.e.
$\tau\sqrt{k\zeta}\rightarrow\infty$, or for negligible acceleration as compared to radiative losses $k/\zeta\rightarrow\infty$ 
the particle distribution reads
\be
f(p)=\frac{1}{a^3}e^{-ap}(2+2ap+a^2p^2)
\vert\zeta-kp^2\vert^{-1}p^{-2}
\label{12}.\ee
For a power law source distribution, $S(p)=s_0p^{-z}$ $f_0$, can be determined
as
\be
f_0=s_0\int^p{p^{\prime 2-z}
e^{-ap^\prime}\left\vert\frac{1+p^\prime\sqrt{k/\zeta}}{1-p^\prime
\sqrt{k/\zeta}}
\right\vert^{\frac{1}{2\tau\sqrt{k\zeta}}}
dp^\prime}.
\label{13}\ee
If one considers again
the limits of negligible particle escape or negligible acceleration
one finds
\be
f(p)=s_0\frac{p^{3-z}}{3-z}\vert\zeta-kp^2\vert^{-1}p^{-2}
\label{14}.\ee

Non-stationary solutions of equation(\ref{7})
with an exponential time evolution $f(p,t)\sim e^{-t/\tau}$ can be found 
provided the distribution function satisfies ${\cal L}_p f = S(p)$.
For a monoenergetic source population one obtains for $p>p_0$
\be
f(p,t)=cp^{-2}\vert\zeta-k p^2\vert^{-1}e^{-\frac{t}{\tau}}
\label{15}\ee
where the constant $c$ contains $s_0$.
A relativistic Maxwellian source population and a power-law one
result in
\be
f(p,t)=\left(c+\frac{s_0}{a}e^{-ap}\right)
p^{-2}\vert\zeta -kp^2\vert^{-1}e^{-\frac{t}{\tau}}
\label{16}\ee
and
\be
f(p,t)=\left(c+s_0p^{-z}\right)
p^{-2}\vert\zeta -kp^2\vert^{-1}e^{-\frac{t}{\tau}}
\label{17},\ee
respectively.

\section{Synchrotron emission spectrum}

The spectral power $P_{\nu}$ radiated by a single particle with momentum $p$
is in a good approximation given by 
($b=\sqrt{3}e^3B\sin{\theta}/mc^2$ and
$\nu_c=3eB\sin{\theta}\thinspace p^2/4\pi(mc)^3$, with $\theta$ being the angle
between the magnetic field and the line of sight)
\be
P_{\nu}(p)=b\left({\frac{\nu}{\nu_c}}\right)^{1/3}e^{-\nu/\nu_c}
\label{18},\ee
see \citet{mel80b}.
It has its maximum near $\nu=\nu_c$ and shows a characteristic $\nu^{1/3}$
spectrum for $\nu\ll\nu_c$ and an exponential cutoff for $\nu\gg\nu_c$.
The specific intensity
from an isotropic momentum distribution of particle density
$N(p)=4\pi p^2f(p)$ can then be calculated by
\be
I_{\nu}=\int dp\ N(p)P_{\nu}(p)
\label{19}.
\ee
In figure~(1) we plot  the distribution function $N(p)=4\pi p^2f(p)$ with
$f(p)$ from equation(\ref{9}) for a monoenergetic source population 
(equation(\ref{10})) with different values of
$a=1/{2\tau\sqrt{k\zeta}}$. The pile-up results for $a=0.5$. 
The higher cut-off results for $a=2$ and the third curve represents 
the case $a=6$.
Figure~(2) shows the corresponding synchrotron
radiation spectra along with the spectrum of a single particle
(upper curve) according to equation(\ref{18}) for comparison. 
Going downward, the other
curves again represent $a=0.5$, $a=2$, and $a=6$, respectively. It is
interesting that the low-frequency part becomes flattened even though
the particle distribution remains constant at low energies and changes
only near the cutoff momentum. The spectral index $\alpha$ with
$I_{\nu}\propto\nu^{-\alpha}$ is given by $\alpha=-0.33$ for the upper curve
and $\alpha\approx -0.1$ for the lowest curve. 
This shows that synchrotron emission with inverted spectra is to be 
expected from reconnection zones. We note again that 
in the context of AGNs and BL Lacs the complete emission
spectra show power-laws for higher than radio frequencies.

\section{Discussion}

Particle acceleration by field-aligned electric fields in
reconnection regions can be described by diffusion in momentum
space as long as the differential momentum changes are small compared to the
respective instant particle momentum. This assumption seems not to be
that restrictive, since reconnection events are expected to excite 
intense Alfv\'enic modes and MHD turbulence that causes efficiently pitch 
angle diffusion.
Thus, in many applications the diffusion treatment seems to be appropriate.
The present analysis shows that mainly constant
distributions for accelerated particles are to be expected no matter whether
monoenergetic, Maxwellian or power law populations are injected
into the reconnection regions. The associate synchrotron spectra
$I_{\nu}\propto\nu^{-\alpha}$ show negative spectral indices $\alpha <0$.
Such rising spectra are ubiquitously observed in galactic nuclei
\citep{ray00} as in the core regions of blazars \citep{edw00}.

We stress that the actual acceleration problem is highly 
involved.  Our findings come from a quite simple nonetheless
physically reasonable scenario. 
Highly anisotropic particle distributions
cannot be described within our approach. Neither local variations of the
spectra can be determined. Whereas some improvement may be possible on the 
analytical side we feel that a numerical test particle approach and
self-consistent electromagnetic particle simulations are indispensable 
for a deeper understanding of the acceleration of high-energy particles
in reconnection regions.

\acknowledgments

This work has been supported by the Deutsche Forschungsgemeinschaft through
the grant LE 1039/6.

\figcaption[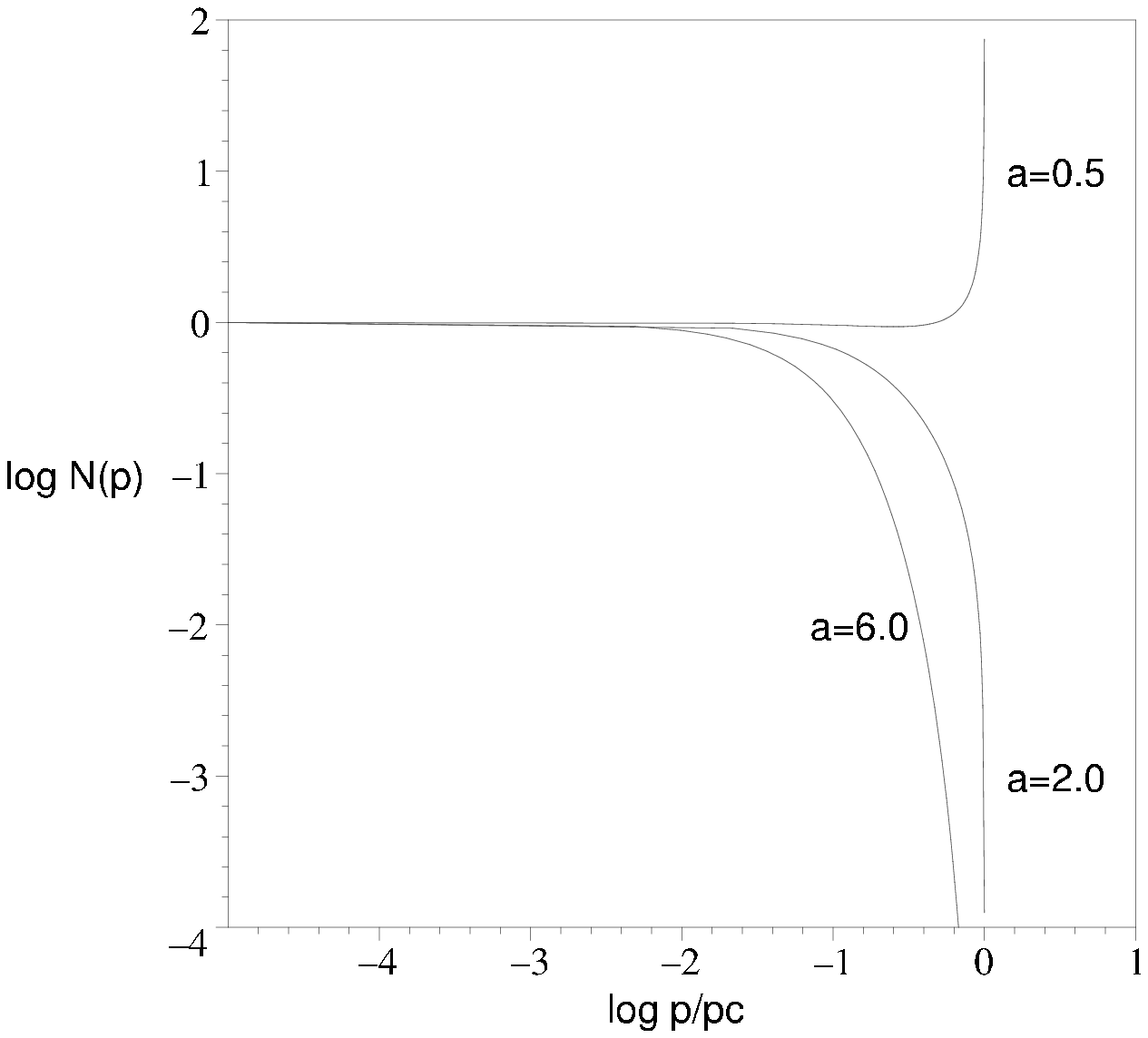]{The particle spectrum in a double-logarithmic
representation for $a=1/{2\tau\sqrt{k\zeta}}$ (see equation(\ref{9}) chosen
as $a=0.5$ (pile-up), $a=2$ (higher cut-off), and $a=6$ (lower cut-off),
respectively. \label{fig1}}

\figcaption[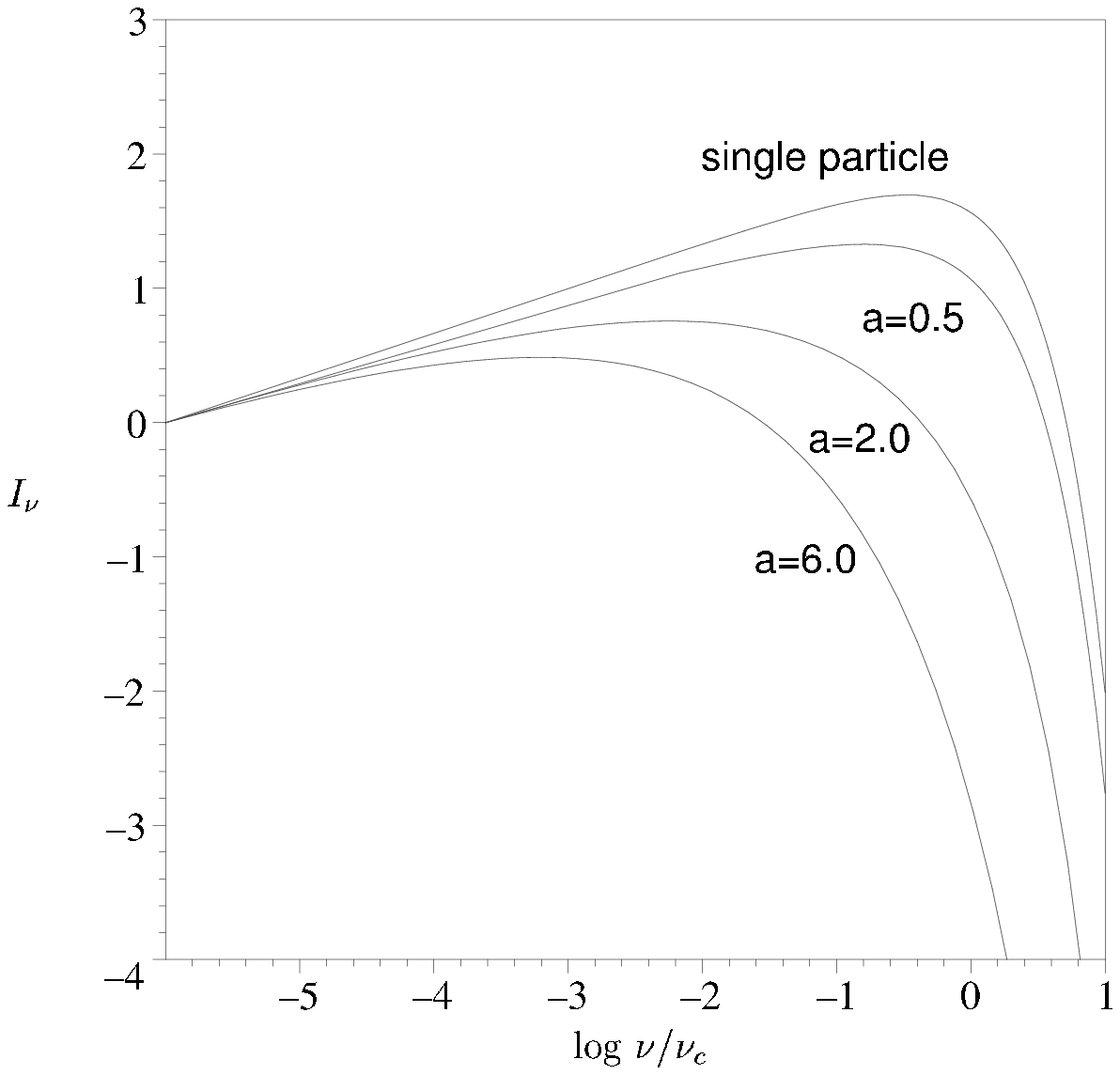]{The synchrotron emission spectrum in a double-logarithmic
representation.
The uppermost curve represents the single-particle spectrum
and the following ones represent the cases $a=0.5$, $a=2$, and $a=6$,
respectively. The logarithm of the intensity is normalized to its value at
${\rm log}(\nu/\nu_c)=-6$.
\label{fig2}}

\plotone{f1.eps}

\plotone{f2.eps}

\end{document}